\documentstyle[12pt]{article}

\begin{document}

\renewcommand{\a}{\alpha}
\renewcommand{\b}{\beta}
\newcommand{\g}{\gamma}           \newcommand{\G}{\Gamma}
\renewcommand{\d}{\delta}         \newcommand{\D}{\Delta}
\newcommand{\ve}{\varepsilon}
\newcommand{\eps}{\epsilon}
\newcommand{\k}{\kappa}
\newcommand{\ld}{\lambda}        \newcommand{\LD}{\Lambda}
\newcommand{\om}{\omega}         \newcommand{\OM}{\Omega}
\newcommand{\p}{\psi}             \newcommand{\PS}{\Psi}
\newcommand{\ro}{\rho}
\newcommand{\s}{\sigma}           \renewcommand{\S}{\Sigma}
\newcommand{\th}{\theta}         \newcommand{\T}{\Theta}
\newcommand{\f}{{\phi}}           \newcommand{\F}{{\Phi}}
\newcommand{\vf}{{\varphi}}
\newcommand{\y}{{\upsilon}}       \newcommand{\Y}{{\Upsilon}}
\newcommand{\z}{\zeta}
\newcommand{\X}{\Xi}
\newcommand{\cA}{{\cal A}}
\newcommand{\cB}{{\cal B}}
\newcommand{\cC}{{\cal C}}
\newcommand{\cD}{{\cal D}}
\newcommand{\cE}{{\cal E}}
\newcommand{\cF}{{\cal F}}
\newcommand{\cG}{{\cal G}}
\newcommand{\cH}{{\cal H}}
\newcommand{\cI}{{\cal I}}
\newcommand{\cJ}{{\cal J}}
\newcommand{\cK}{{\cal K}}
\newcommand{\cL}{{\cal L}}
\newcommand{\cM}{{\cal M}}
\newcommand{\cN}{{\cal N}}
\newcommand{\cO}{{\cal O}}
\newcommand{\cP}{{\cal P}}
\newcommand{\cQ}{{\cal Q}}
\newcommand{\cS}{{\cal S}}
\newcommand{\cR}{{\cal R}}
\newcommand{\cT}{{\cal T}}
\newcommand{\cU}{{\cal U}}
\newcommand{\cV}{{\cal V}}
\newcommand{\cW}{{\cal W}}
\newcommand{\cX}{{\cal X}}
\newcommand{\cY}{{\cal Y}}
\newcommand{\cZ}{{\cal Z}}
\newcommand{\hA}{{\widehat A}}
\newcommand{\hB}{{\widehat B}}
\newcommand{\hC}{{\widehat C}}
\newcommand{\hD}{{\widehat D}}
\newcommand{\hE}{{\widehat E}}
\newcommand{\hF}{{\widehat F}}
\newcommand{\hG}{{\widehat G}}
\newcommand{\hH}{{\widehat H}}
\newcommand{\hI}{{\widehat I}}
\newcommand{\hJ}{{\widehat J}}
\newcommand{\hK}{{\widehat K}}
\newcommand{\hL}{{\widehat L}}
\newcommand{\hM}{{\widehat M}}
\newcommand{\hN}{{\widehat N}}
\newcommand{\hO}{{\widehat O}}
\newcommand{\hP}{{\widehat P}}
\newcommand{\hQ}{{\widehat Q}}
\newcommand{\hS}{{\widehat S}}
\newcommand{\hR}{{\widehat R}}
\newcommand{\hT}{{\widehat T}}
\newcommand{\hU}{{\widehat U}}
\newcommand{\hV}{{\widehat V}}
\newcommand{\hW}{{\widehat W}}
\newcommand{\hX}{{\widehat X}}
\newcommand{\hY}{{\widehat Y}}
\newcommand{\hZ}{{\widehat Z}}
\newcommand{\Ha}{{\widehat a}}
\newcommand{\Hb}{{\widehat b}}
\newcommand{\Hc}{{\widehat c}}
\newcommand{\Hd}{{\widehat d}}
\newcommand{\He}{{\widehat e}}
\newcommand{\Hf}{{\widehat f}}
\newcommand{\Hg}{{\widehat g}}
\newcommand{\Hh}{{\widehat h}}
\newcommand{\Hi}{{\widehat i}}
\newcommand{\Hj}{{\widehat j}}
\newcommand{\Hk}{{\widehat k}}
\newcommand{\Hl}{{\widehat l}}
\newcommand{\Hm}{{\widehat m}}
\newcommand{\Hn}{{\widehat n}}
\newcommand{\Ho}{{\widehat o}}
\newcommand{\Hp}{{\widehat p}}
\newcommand{\Hq}{{\widehat q}}
\newcommand{\Hs}{{\widehat s}}
\newcommand{\Hr}{{\widehat r}}
\newcommand{\Ht}{{\widehat t}}
\newcommand{\Hu}{{\widehat u}}
\newcommand{\Hv}{{\widehat v}}
\newcommand{\Hw}{{\widehat w}}
\newcommand{\Hx}{{\widehat x}}
\newcommand{\Hy}{{\widehat y}}
\newcommand{\Hz}{{\widehat z}}
\newcommand{\deff}{\,\stackrel{\rm def}{\equiv}\,}
\newcommand{\lra}{\longrightarrow}
\newcommand{\ra}{\,\rightarrow\,}
\def\limar#1#2{\,\raise0.3ex\hbox{$\longrightarrow$\kern-1.5em\raise-1.1ex
\hbox{$\scriptstyle{#1\rightarrow #2}$}}\,}
\def\limarr#1#2{\,\raise0.3ex\hbox{$\longrightarrow$\kern-1.5em\raise-1.3ex
\hbox{$\scriptstyle{#1\rightarrow #2}$}}\,}
\def\limlar#1#2{\ \raise0.3ex
\hbox{$-\hspace{-0.5em}-\hspace{-0.5em}-\hspace{-0.5em}
\longrightarrow$\kern-2.7em\raise-1.1ex
\hbox{$\scriptstyle{#1\rightarrow #2}$}}\ \ }
\newcommand{\limm}[2]{\lim_{\stackrel{\scriptstyle #1}{\scriptstyle #2}}}
\newcommand{\wt}{\widetilde}
\newcommand{\os}{{\otimes}}
\newcommand{\da}{{\dagger}}
\newcommand{\stimes}{\times\hspace{-1.1 em}\supset}
\def\h{\hbar}
\newcommand{\ih}{\frac{\i}{\h}}
\newcommand{\exx}[1]{\exp\left\{ {#1}\right\}}
\newcommand{\ord}[1]{\mbox{\boldmath{$\cO$}}\left({#1}\right)}
\newcommand{\one}{{\leavevmode{\rm 1\mkern -5.4mu I}}}
\newcommand{\Z}{Z\!\!\!Z}
%
\newcommand{\Ibb}[1]{ {\rm I\ifmmode\mkern
            -3.6mu\else\kern -.2em\fi#1}}
\newcommand{\ibb}[1]{\leavevmode\hbox{\kern.3em\vrule
     height 1.2ex depth -.3ex width .2pt\kern-.3em\rm#1}}
\newcommand{\N}{{\Ibb N}}
\newcommand{\C}{{\ibb C}}
\newcommand{\R}{{\Ibb R}}
\newcommand{\HH}{{\Ibb H}}
\newcommand{\rational}{{\kern .1em {\raise .47ex
\hbox{$\scripscriptstyle |$}}
    \kern -.35em {\rm Q}}}
\newcommand{\bm}[1]{\mbox{\boldmath${#1}$}}
\newcommand{\intf}{\int_{-\infty}^{\infty}\,}
\newcommand{\LL}{\cL^2(\R^2)}
\newcommand{\LLS}{\cL^2(S)}
\newcommand{\Ree}{{\cal R}\!e \,}
\newcommand{\Imm}{{\cal I}\!m \,}
\newcommand{\tr}{{\rm {Tr} \,}}
\newcommand{\er}{{\rm{e}}}
\renewcommand{\i}{{\rm{i}}}
\newcommand{\divv}{{\rm {div} \,}}
\newcommand{\id}{{\rm{id}\,}}
\newcommand{\ad}{{\rm{ad}\,}}
\newcommand{\Ad}{{\rm{Ad}\,}}
\newcommand{\const}{{\rm{\,const\,}}}
\newcommand{\rank}{{\rm{\,rank\,}}}
\newcommand{\diag}{{\rm{\,diag\,}}}
\newcommand{\sign}{{\rm{\,sign\,}}}
\newcommand{\pa}{\partial}
\newcommand{\pad}[2]{{\frac{\partial #1}{\partial #2}}}
\newcommand{\padd}[2]{{\frac{\partial^2 #1}{\partial {#2}^2}}}
\newcommand{\paddd}[3]{{\frac{\partial^2 #1}{\partial {#2}\partial {#3}}}}
\newcommand{\der}[2]{{\frac{{\rm d} #1}{{\rm d} #2}}}
\newcommand{\derr}[2]{{\frac{{\rm d}^2 #1}{{\rm d} {#2}^2}}}
\newcommand{\fud}[2]{{\frac{\delta #1}{\delta #2}}}
\newcommand{\fudd}[2]{{\frac{\d^2 #1}{\d {#2}^2}}}
\newcommand{\fuddd}[3]{{\frac{\d^2 #1}{\d {#2}\d {#3}}}}
\newcommand{\dpad}[2]{{\displaystyle{\frac{\partial #1}{\partial #2}}}}
\newcommand{\dfud}[2]{{\displaystyle{\frac{\delta #1}{\delta #2}}}}
\newcommand{\dd}{\partial^{(\ve)}}
\newcommand{\ddd}{\bar{\partial}^{(\ve)}}
\newcommand{\dfrac}[2]{{\displaystyle{\frac{#1}{#2}}}}
\newcommand{\dsum}[2]{\displaystyle{\sum_{#1}^{#2}}}
\newcommand{\dint}{\displaystyle{\int}}
\newcommand{\dg}{\!\not\!\partial}
\newcommand{\vg}[1]{\!\not\!#1}
\def\<{\langle}
\def\>{\rangle}
\def\lgl{\langle\langle}
\def\rgr{\rangle\rangle}
\newcommand{\bra}[1]{\left\langle {#1}\right|}
\newcommand{\ket}[1]{\left| {#1}\right\rangle}
\newcommand{\vev}[1]{\left\langle {#1}\right\rangle}
\newcommand{\be}{\begin{equation}}
\newcommand{\ee}{\end{equation}}
\newcommand{\bn}{\begin{eqnarray}}
\newcommand{\en}{\end{eqnarray}}
\newcommand{\bnn}{\begin{eqnarray*}}
\newcommand{\enn}{\end{eqnarray*}}
\newcommand{\e}{\label}
\newcommand{\nbr}{\nonumber\\[2mm]}
\newcommand{\r}[1]{(\ref{#1})}
\newcommand{\refp}[1]{\ref{#1}, page~\pageref{#1}}
\renewcommand {\theequation}{\thesection.\arabic{equation}}
\renewcommand {\thefootnote}{\fnsymbol{footnote}}
\newcommand{\qq}{\qquad}
\newcommand{\qqq}{\quad\quad}
\newcommand{\biz}{\begin{itemize}}
\newcommand{\eiz}{\end{itemize}}
\newcommand{\ben}{\begin{enumerate}}
\newcommand{\een}{\end{enumerate}}
\def\nc{noncommutative }
\def\ncy{noncommutativity }
\def\com{commutative }
\def \simlt{\stackrel{<}{{}_\sim}}
\thispagestyle{empty}
\begin{flushright}
HIP-2001-01/TH\\
hep-th/0012175
\end{flushright}

\begin{center}

{\Large{\bf{Aharonov-Bohm Effect in Noncommutative Spaces}}}
\vskip .7cm
{\bf{\large{M. Chaichian$^{\dagger}$, A. Demichev$^{\dagger,a}$
P. Pre\v{s}najder$^{\dagger,b}$, M. M. Sheikh-Jabbari$^{\dagger\dagger}$
\ \ and \ \ {{A. Tureanu}}$^{\dagger}$}}}

{\it $^{\dagger}$High Energy Physics Division, Department of
Physics,
University of Helsinki\\
\ \ {and}\\
\ \ Helsinki Institute of Physics,
P.O. Box 64, FIN-00014 Helsinki, Finland\\
$^a$Nuclear Physics Institute, Moscow State University, 119899 Moscow,
Russia \\ 
$^b$Department of Theoretical Physics, Comenius University, Mlynsk\'{a}
dolina, SK-84248 Bratislava,
Slovakia \\
$^{\dagger\dagger}$ The Abdus Salam International Center for Theoretical
Physics,\\
Strada Costiera 11,Trieste, Italy}

\setcounter{footnote}{0}

{\bf abstract}

\end{center}  

The Aharonov-Bohm effect on the \nc plane is considered. Developing the path integral 
formulation of quantum mechanics, we find the propagation amplitude for a particle in a \nc
space. We show that the 
corresponding shift in the phase of the particle propagator due to the 
magnetic field of a thin solenoid receives certain gauge invariant
corrections because of the noncommutativity. Evaluating the numerical value for this correction, an upper
bound for the noncommutativity parameter is obtained.


\newpage
\section{Introduction}
\setcounter{equation}{0}

Besides the string theory interests \cite{SW}, recently theories on \nc
space-time have received a lot of 
attention. In this way both problems of quantum mechanics (QM) and field theories on
\nc spaces have their own
excitements; for the QM side see \cite{{Sny},{Alvarez},{Lamb}}. The \nc
spaces can be
realized as spaces
where coordinate operators, $\hat{x}^{\mu}$, satisfy the commutation
relations
\be\label{cr}
[\hat{x}^{\mu},\hat{x}^{\nu}]=\i\theta^{\mu\nu}\ , 
\ee 
where $\theta_{\mu\nu}$ is an antisymmetric tensor of dimension of
(length)$^2$.
We note that  
a space-time noncommutativity, $\theta_{0i}\neq 0$,
may lead to some problems with unitarity and causality
\cite{{GoM},{CDP3}}.
Such problems do not occur for the QM on a \nc space with a usual
(commutative) time coordinate. 

Given the \nc space (\ref{cr}), which is also a natural extension of the usual
QM, one should study its physical consequences. Comparing these \nc
results with the present experimental data one can find an upper bound on 
$\theta$.
It appears that the most natural places to trace the \ncy effects are
simple QM systems, such as the hydrogen atom or the Aharonov-Bohm effect. The
former has been considered in \cite{Lamb} and the shift in the   
spectrum, and in particular the modifications to Lamb-shift, due to \ncy
have been discussed there. As one expects (\ref{cr}) breaks the rotational
symmetry of the hydrogen atom spectrum and as a result we would face a
"polarized Lamb-shift" \cite{Lamb}. In this work we shall study
the other system, the Aharonov-Bohm effect. The physical significance of the 
Aharonov-Bohm effect resides in the fact that it is the place to check the
\nc gauge invariance, which is a deformed version of the usual gauge
freedom \cite{{SW},{Alvarez},{Ihab}}. We will come back to this point
later. 

In order to study the Aharonov-Bohm effect one should develop the proper
QM setup for the \nc case.
As the Hilbert space is assumed to be the same for the \com 
case and its \nc extension, it is enough to give the Hamiltonian. 
Once we have the Hamiltonian, the dynamics of the states is given by the
usual Schroedinger equation,
$H|\psi\rangle =i\hbar{\partial \over \partial t}|\psi\rangle$. 
Because of the noncommutativity of the coordinates, the coordinate basis does not exist and
the very concept of wave function, $\langle x|\psi\rangle$, fails.
However, the usual momentum space description is still valid. 

To handle the NCQM, one can also use a more unusual approach to QM, using the
operator valued "wave functions". In the usual QM because of the Weyl-Moyal
correspondence \cite{{Alvarez},{Ihab}} there is a one-to-one correspondence between 
such operators and the usual wave functions so that the usual algebra of the
functions is now applicable to them. However, in the \nc case, instead of the usual
product between functions, the Weyl-Moyal correspondence yields the $\star$-product:
\begin{eqnarray}\label{star}
(f\star g)(x)&=&\exx{{\i\over 2}\theta_{\mu\nu}
\partial_{x_{\mu}}\partial_{y_{\nu}}}f(x)g(y)\Big|_{x=y}\cr
&=&f(x)g(x)+{\i\over2}\theta_{\mu\nu}
\partial_{{\mu}}f\partial_{{\nu}}g+\ord{\theta^2}, 
\end{eqnarray}
between the "wave functions". 
We should remind that, although always a function
corresponds  to any operator valued wave function, the argument of these
functions cannot be treated as the space coordinates. According to this
point of
view the probability amplitudes 
are given by the square of the norm of the operator valued wave functions.

Having discussed the NCQM kinematics, we should then give the proper Hamiltonian for the 
\nc systems. As in the usual QM, this can be done using the
non-relativistic limit of the 
corresponding field theory. The difference between the \com and \nc field theories
are only in the interaction terms \footnote{Of course this is not quite
true, and for the field theories on \nc spaces with non-trivial topology,
such as cylinder and torus, one
should treat the problem more carefully \cite{Appear}.}
\cite{{Ihab},{Andrei}}, and this
will lead to some new $\theta$ dependent interaction potentials.   
For the electro-magnetic interactions the corresponding field theory is NCQED. As
discussed in \cite{{Ihab},{Lamb}} for the electro-magnetic interaction
the extra $\theta$ dependence of the Hamiltonian, in the first order in
$\theta$, always can be obtained assigning an electric dipole moment,
\be\label{dip}
d_e^i={e\over 2\hbar}\theta^{ij}p_j\ ,
\ee
to the charged particle.
This can be understood intuitively noting that $f(x)\star g(x)=f(x_i+{i\over
2}\theta_{ij}\partial_j)g(x)$.
Since we believe that the effect of \ncy in nature, if it is there, should be very small,
one can trust the perturbation in $\theta$. 

It turns out that to study the Aharonov-Bohm effect it is more
convenient to formulate the problem via path integral. So, first we
construct the proper definition of the path integral and transition
amplitude in the \nc case and then we
evaluate the extra shift in the Aharonov-Bohm experiment interference
pattern which comes about due to noncommutativity in the {\it quasi 
classical} approximation, i.e. leading order in $\hbar$ and first order in
$\theta$.

\section {Aharonov-Bohm effect on a noncommutative plane\e{BAep}}
\setcounter{equation}{0}

The Aharonov-Bohm effect concerns the shift of the interference pattern in the double-slit experiment, due to
the presence of a thin long solenoid put just between the two slits
\cite{Feynman}. 
Although the magnetic field $B$  is present only inside the solenoid, the
corresponding Schroedinger equation
depends explicitly on the magnetic potential $A$ (non-vanishing outside
the solenoid). Therefore, the wave
function depends on $A$ and consequently the interference pattern shifts. 
The shift in the phase of the particles propagator, $\delta \phi_0$, is
gauge invariant itself and can be
expressed in non-local terms of $B$. In the quasi-classical
approximation, $\delta\phi_0={e\over \hbar c}\Phi$,
where $\Phi=B\pi\rho^2$ is the magnetic flux
through the solenoid of radius $\rho$. This effect has been confirmed
experimentally \cite{Tono}.

Below, we present the quasi-classical approach to the Aharonov-Bohm
effect on a
NC-plane for a thin, but of finite radius solenoid. However first we need
to formulate a \nc path integral.

Since the very concept of the wave function in the \nc case is a
problematic one, in order to study
the \nc Aharonov-Bohm effect, first we present the \nc formulation of
path integral QM. Then, by means of path integrals, 
we find the propagator and hence the desired \nc corrections to  the Aharonov-Bohm phase. 

The Hilbert space $\cH$ of quantum mechanics on a \nc space is formed by the normalizable
functions
$\Psi(x)$ with finite norm, belonging to a \nc algebra of functions $\cA$ on $\R^2$. The wave function is an element from $\cH$, normalized to unity.
However, we remind that wave functions are just symbols; the physical
meaning is contained only in some smeared values of
them, e.g. by a coherent state \cite{CDP1}.
In $\cA$, we can introduce the scalar product as:
\begin{eqnarray}
(\psi,\phi)&=&\int d^3x\bar{\psi}(x)\star \phi(x)=\int d^3x\bar{\psi}(x)\phi(x)\cr
&=&\int d^3k\bar{\tilde{\psi}}(k)\tilde{\phi}(k)\ ,
\end{eqnarray}
where $\tilde\psi (k)$ and $\tilde\phi (k)$ are the corresponding Fourier
transforms. 
Here, we have used the well-known fact that in the integrals containing as integrand a $\star$-product of two
functions, their $\star$-product can be replaced by a standard one.
The operators $P_i$ and $X_i$ acting in $\cH$ and satisfying Heisenberg canonical commutation relations are
defined by:
\be\label{op}
P_i\Psi(x)=-\i\partial_i\Psi(x)\ ,\qq  X_i\Psi(x)=x_i\star \Psi(x)\ .
\ee
Along the arguments of \cite{Lamb}, the problem of a particle moving in an external magnetic field on a \nc
plane is specified by the Hamiltonian:
\be\label{Hamil}
H={1\over 2}(P_i+A_i)^2\ .
\ee
We note that the transition amplitude $(\Psi_f,\er^{-\i H t}\Psi_i)$ is invariant under the \nc gauge 
transformations defined by
\begin{eqnarray}
\Psi(x)&\rightarrow& U(x)\star\Psi(x)\ , \cr
A_i(x)&\rightarrow& U(x)\star A_i(x)\star U^{-1}(x) 
-\i U(x)\star \partial_iU^{-1}(x)\ ,
\nonumber
\end{eqnarray}
where  $U(x)\equiv(\er\star)^{i\lambda(x)}$, for real functions
$\lambda(x)$, and the $(\er\star)$ is defined
by the usual Taylor expansion, with all  products of $\lambda$'s 
replaced by the $\star$ ones.
Then, one can easily show that $U^{-1}=(\er\star)^{-\i\lambda(x)}$ satisfies $U^{-1}\star U=1$.
We point out the non-Abelian character of the above gauge transformations, due to the noncommutativity of the
space. Consequently, the field strength is given by a non-Abelian formula, too:
\be
F_{ij}(x)=\partial_{[i}A_{j]}(x)+(A_{[i}\star
A_{j]})(x)\ .
\ee
Moreover, one can easily see that
\be\label{trgauge}
P_i+A_i\rightarrow U(x)\star(P_i+A_i)\star U^{-1}(x)\ .
\ee

In quantum mechanics, the exponents of the operators (e.g., $\er^{-\i H t}$) often do not correspond to local
operators. However, they can be conveniently represented by bi-local kernels. This is true in the \nc frame,
also. It can be easily seen that to any operator $K=K(P_i,X_i)=K(-\i\partial_i,x_i\star)$,
cf.(\ref{op}), we can assign a kernel (a bi-local symbol)
$\cK(x,y)\in\cA\otimes\cA$, defined by:
\be
\cK(x,y)=\int {d^3q\over (2\pi)^3} (K\er^{\i q x})\er^{-\i q y}.
\ee
(we omit the symbol $\otimes$ for the direct product). 
We note that straightforwardly the $\star$-product defined between two
functions (Weyl symbols) can be generalized to the kernels which are
functions in two variables. The action of $K$ in terms of kernel is
$$
(K\Phi)(x)=\int d^3y\cK(x,y)\star\Phi(y)=\int d^3y\cK(x,y)\Phi(y)\ .
$$
For a product of two operators, one can use either the standard formula
for the kernel composition
\begin{eqnarray}\label{prodop1}
(\cG\cK)(x,y)&=&\int d^3z\cG(x,z)\star\cK(z,y)\cr
&=&\int d^3z\cG(x,z)\cK(z,y)\ .
\end{eqnarray}
or use the formula
\be\label{prodop2}
(\cG\cK)(x,y)=\int{d^3q\over (2\pi)^3}(G\er^{\i q
x})\overline{(K^{\dagger}\er^{\i q y})}\ .
\ee
The proof of (\ref{prodop2}) is straightforward.

The kernel corresponding to the operator $\er^{-\i H t}$ will be denoted by $\cK_t(x,y)$ and called
propagator:
\be\label{Kt}
\cK_t(x,y)=\int{d^3q\over (2\pi)^3}(\er^{-\i H t}\er^{\i q x})\er^{-\i q y}\ .
\ee
From the product formula (\ref{prodop1}) and the identity $\er^{-\i H t_1}\er^{-\i H t_2}=\er^{-\i H
(t_1+t_2)}$, the usual composition law follows:
\be
\cK_{t_1+t_2}(x,y)=\int d^3z\cK_{t_1}(x,z)\cK_{t_{2}}(z,y)\ .
\ee
Iterating this formula $N$ times and taking the limit $N\rightarrow\infty$, we arrive by standard arguments at
the path integral representation of the propagator:
\begin{eqnarray}\label{propagator}
\cK_t(x,y)&=&\lim_{N\rightarrow\infty}\int d^3x_{N-1}\cdots
d^3x_1 \cr
&\times&\cK_{\epsilon}(x,x_{N-1})\cdots\cK_{\epsilon}(x_2,x_1)\cK_{\epsilon}(x_1,y)\ ,
\end{eqnarray}
with $\epsilon=t/N$. We stress that there is no need to use $\star$-product between two $\cK_{\epsilon}$'s. 

The formula for the gauge transformation of the propagator follows directly from  eq. (\ref{trgauge}). 
In fact, (\ref{trgauge}) implies:
\be
\er^{-\i H t}\rightarrow U(x)\star\er^{-\i H t}\star U^{-1}(x)\ ,
\ee
(as operators), so that
\begin{eqnarray}
\cK_t(x,y)&\rightarrow& U(x)\star\er^{-\i H t}\star U^{-1}(y)\cr
&=&\int {d^3q\over (2\pi)^3}\bigl(\er^{\i \lambda}\star(\er^{-\i H t}\er^{\i q
x})\bigr)\star(\overline{\er^{\i
\lambda}\er^{\i q y}})\cr
&=&U(x)\star\cK_t(x,y)\star U^{-1}(y)\ .
\end{eqnarray}
This is exactly the expected formula (here, the $\star$-product cannot be omitted).

As the next step, we shall calculate the short-time propagator $\cK_{\epsilon}(x,y)$ entering
(\ref{propagator}) to first orders in $\epsilon$ and $\theta$. Using the Hamiltonian (\ref{Hamil}) and
(\ref{Kt}) we have
\begin{eqnarray}
\cK_{\epsilon}(x,y)&=&\int{d^3p\over (2\pi)^3}\biggl([1-{\i\epsilon\over
2}(P_i+A_i)^2+\cdots]\er^{\i p x}\biggr)\er^{-\i p y}\cr\nonumber
&=&\int{d^3p\over (2\pi)^3}\er^{\i p (x-y)-{\i\epsilon}H_e(p,\bar{x})}\ ,
\nonumber
\end{eqnarray}
where $\bar{x}={1\over 2}(x+y)$ and the effective Hamiltonian, $H_e$, is given as
\footnote{We note that $A\star A=A^2+O(\theta^2)$.}:
\be
H_e\cong{1\over 2}(\Pi_i+A_i(\bar{x}))^2\, ;\Pi_i=p_i-{1\over 2}\theta_{jk}(\partial_{j}A_i(\bar{x}))p_{k}.
\ee
The symbol $\cong$ means equality in the first order in $\epsilon$ and
$\theta$. 
The above effective Hamiltonian 
can also be obtained if we assign an electric dipole moment,
eq.~(\ref{dip}), to electron. 
Performing the $d^3p$
integration, we obtain the effective Lagrangian:
\be\label{efflagr}
\cL\cong{1\over 2}V_i V_i-V_i A_i(\bar{x})\, ; \ V_i=v_i+{1\over
2}\theta_{ji}\partial_{j}A_k(\bar{x})v_{k}.
\ee
The formula for $\cK_{\epsilon}(x,y)$ then reads:
\be
\cK_{\epsilon}(x,y)\cong\er^{\i\int dt\cL(\bar{x}(t),\dot{\bar{x}}(t))}\ ,
\ee
with the effective action calculated for a linear path, starting at $x_i(0)=x_i$ and terminating at
$x_i(\epsilon)=y_i$, i.e., $v_i=(y_i-x_i)/\epsilon$ and $A_i(\bar{x})=A_i({x+y\over 2})$. Up to terms linear
in $\theta$, the Lagrangian, with all physical constants included, becomes:
\begin{eqnarray}\label{lagr}
\cL=\cL_0-{e m\over 4\hbar
c}\vec{\theta}\cdot\bigl[v_i(\vec{v}\times\vec{\nabla}A_i)-
{e\over mc}v_i(\vec{A}\times\vec{\nabla}A_i)\bigr],
\end{eqnarray}
where $\cL_0={m\over 2}\vec{v}^2-{e\over c}\vec{v}\cdot\vec{A}$ and
the vector $\vec\theta$ is defined as
$\theta_i=\epsilon_{ijk}\theta_{jk}$.
Thus, the total shift of phase for the Aharonov-Bohm effect, including the contribution due to
noncommutativity, will be:
\be\label{ncshift}
\delta\phi_{total}=\delta\phi_0+\delta\phi_{\theta}^{NC}\ ,
\ee
where $\delta\phi_0={e\over \hbar c}\oint d\vec{r}\cdot\vec{A}=
{e\over \hbar c}\int \vec{B}\cdot d\vec{S}={e\over \hbar c}\Phi$ ($\Phi$ being
the magnetic flux through the surface bounded by the closed path)
is the usual (commutative) phase shift and
\be\label{nccorr}
\delta\phi_{\theta}^{NC}={e m\over 4\hbar^2 c}\vec{\theta}\cdot
\oint dx_i\bigl[(\vec{v}\times\vec{\nabla}A_i)-
{e\over mc}(\vec{A}\times\vec{\nabla}A_i)\bigr]
\ee
represents the \nc corrections. 
 
For a finite-radius solenoid, the vector potential $\vec{A}$ entering (\ref{lagr})-(\ref{nccorr}) is given by:
\be\label{solennoid}
\vec{A}={1\over 2}B{\rho^2\over r}\vec{n}\ ,\qq r>\rho\ ,
\ee
where $B$ is the constant magnetic field inside the solenoid, $\rho$ is the radius of the solenoid and $\vec{n}$
is the unit vector orthogonal to $\vec{r}$.

The expression for the correction $\delta\phi_{\theta}^{NC}$ to the usual Aharonov-Bohm phase 
due to noncommutativity can be explicitly obtained from (\ref{nccorr}) and (\ref{solennoid}).
In an analogous way as in 
the usual Aharonov-Bohm case \cite{Feynman}, the calculation can be done by taking the closed classical
path (what is valid according to the experimental setup), which starts from the source and reaches
the point on the screen by passing through one of the two slits and returns to the source point through the
other slit.

An estimation for the upper bound on the parameter of noncommutativity $\theta$ can be made using
the available experimental data on the Aharonov-Bohm effect \cite{Tono}. For the purpose of this estimation,
we took $\vec{\theta}$ along the magnetic field of the solenoid, for in this case the effect is the largest; the
integration path was taken to be circular, although this is not significant and the result would be general.
Then, the contribution to the shift coming from noncommutativity, relative to the usual shift of phase, will
be:
\be\label{ratio}
\frac{\delta \phi_{\theta}^{NC}}{\delta\phi_0}\sim
\ \frac{\theta}{\lambda_eR}\frac{v}{c}-\delta\phi_0\frac{\theta}{S}\ ,
\ee
where $R$ is the radius of the approximate path, $S= \pi R^2$ is the area of the surface bounded by the
closed path and $\lambda_e$ is the Compton wavelength of the electron. 
We should point out the fact that, comparing the two terms of the \nc correction, it appears that the
energy-dependent term prevails over the other one (by 5 orders of magnitude). 
Fitting the ratio (\ref{ratio}) into the accuracy
bound of the experiment \cite{Tono}, we obtain:
\be
\sqrt{\theta}\simlt 10^6 GeV^{-1},
\ee
which corresponds to a relatively large scale of 1 Å. Such a value
emerges due to the large error of 20\% in
the experimental test \cite{Tono} of the Aharonov-Bohm effect.
\section{Concluding remarks}

In this work we have studied the Aharonov-Bohm effect for the \nc case. In order to 
obtain the transition amplitudes, and hence finding the shift in the interference
pattern, we worked out the path integral formulation. Using this formulation 
we have also required the transition amplitudes to be
invariant under the \nc gauge transformations. In this way we have found
the result in the {\it quasi-classical}
approach up to first order in $\theta$. However, besides the quasi-classical result one
can solve the Schroedinger equation for this case explicitly, though for the operator
valued wave functions \cite{Appear}.

\vskip 0.3cm
{\Large\bf{Acknowledgements}} 

The financial support of the Academy of Finland under the Project No. 163394
is greatly acknowledged.
A.D.'s work was partially supported by RFBR-00-02-17679 grant
and P.P.'s work by VEGA project 1/7069/20.
The work of M.M.Sh-J was partially supported by the EC contract no. ERBFMRX-CT 96-0090.


\begin{thebibliography}{99}

\bibitem{SW}
N. Seiberg and E. Witten, {\it JHEP} {\bf  9909: 032} (1999).

\bibitem{Sny}
H. Snyder, {\it Phys. Rev.} {\bf 71} (1947) 38.

\bibitem{Alvarez}
L. Alvarez-Gaume, S. R. Wadia, 
Report no. hep-th/0006219.  

\bibitem{Lamb}
M. Chaichian, M. M. Sheikh-Jabbari and A. Tureanu, 
Report no. hep-th/0010175.

\bibitem{GoM}
J. Gomis and T. Mehen, {\it Nucl.Phys.} {\bf B591} (2000) 265.


\bibitem{CDP3}
M. Chaichian, A. Demichev, P. Pre\v{s}najder and A. Tureanu, 
Report no. hep-th/0007156.



\bibitem{Ihab}
I.F. Riad, M.M. Sheikh-Jabbari, {\it JHEP} {\bf 0008:045}. 

\bibitem{Andrei}
A. Micu, M.M. Sheikh-Jabbari, 
{\it JHEP} {\bf 01} (2001) 025,  Report no. hep-th/0008057.

 
\bibitem{CDP1}
M. Chaichian, A. Demichev and P. Pre\v{s}najder, {\it Nucl. Phys.}
{\bf B567} (2000) 360 ; {\it J.Math.Phys.} {\bf 41} (2000) 185.


\bibitem{Feynman}
Richard P. Feynman, Robert B. Leighton and Matthew Sands, {\it The Feynman lectures on physics}, vol. II,
(Reading, MA : Addison-Wesley, 1964).
B. R. Holstein, {\it Topics in advanced quantum mechanics} 
(Addison-Wesley, Redwood City, 1992).


\bibitem {Tono}
A. Tonomura et al., {\it Phys.Rev.Lett.} {\bf 48} (1982) 1443.

\bibitem{Appear}

M. Chaichian, A. Demichev, P. Pre\v{s}najder, M.M. Sheikh-Jabbari,
A. Tureanu, 
Report no. hep-th/0101209.


\end{thebibliography}
\end{document}